\documentclass{article}
\usepackage{spconf,amsmath,graphicx,hyperref}
\usepackage{booktabs}
\usepackage{amssymb}

\title{TargetSEC: Plug-and-Play In-the-Wild Speech Emotion Conversion via Arousal-Conditioned Latent Style Diffusion}
\name{Constantin Alexander Auga}
\address{Hasso Plattner Institute / University of Potsdam,
Potsdam, Germany\\
\texttt{constantin.auga@uni-potsdam.de}}
\begin{document}
\ninept
\maketitle
\begin{abstract}
Speech Emotion Conversion (SEC) aims to transform the emotion of a source utterance into a target emotion while preserving content and speaker identity. SEC on in-the-wild data is challenging due to the non-parallel nature of training data and complex real-world acoustics. Existing fixed-duration approaches either struggle to shift the emotion effectively (high quality, low conversion) or degrade speech naturalness (low quality, high conversion). We propose TargetSEC, an embedding-driven latent diffusion framework that generates emotion-focused style embeddings conditioned on speaker identity and continuous emotion. Unlike methods that diffuse over spectrograms, TargetSEC operates in a compact latent space. Experiments on the MSP-Podcast dataset show that TargetSEC outperforms current non-duration baselines and matches or exceeds the duration-prediction baseline in conversion MSE and naturalness without explicit temporal modeling. Demo and code available at \footnote{\url{https://github.com/conscht/TargetSEC}}.
\end{abstract}
\section{Introduction}
\label{sec:intro}

While state-of-the-art speech synthesis models~\cite{VALE, shen2023naturalspeech2latentdiffusion, li2023styletts2humanleveltexttospeech} excel at producing natural speech, they still struggle to convey nuanced emotions adequately \cite{INOUE202135}, often resulting in robotic or monotonous speech that limits expressive capability \cite{AgeofAIEmo, zhou2023speechsynthesismixedemotions}. Speech Emotion Conversion (SEC) addresses this by explicitly modeling and controlling emotion during synthesis \cite{HMM}.

Recent Voice Conversion (VC) models employ text-prompt-driven Latent Diffusion Models (LDM) to enable stylistic adjustments in audio generation through natural language descriptions \cite{hai2024dreamvoice, promptvcflexiblestylisticvoice}. While these models are capable of performing style conversion and, to some extent, emotion conversion, their inherent capabilities do not allow granular control over the emotional state of the speaker. 

In contrast, models that are specifically designed for SEC represent emotion as categorical or continuous embeddings. While many SEC approaches favor categorical representation due to their simplicity, it is well established in the SEC community and psychology literature that emotions have fuzzy boundaries and cannot be accurately modeled by a categorical representation \cite{MSP, EmoCon}.
The circumplex model addresses this impediment by representing emotions in a continuous space, typically along arousal (activation) and valence (positivity) dimensions. 

However, as the audio modality captures arousal more effectively than valence \cite{navinAR, de_Oliveira_2023}, we follow prior work  \cite{Prabhu2023, EmoCon} and focus on the continuous arousal dimension.

Most existing SEC approaches primarily leverage high-quality acted-out datasets. However, acted-out speech requires professional actors and an extensive recording effort. Moreover, such datasets often exhibit exaggerated emotional expressions and inherent biases. Finally, methods trained on acted-out datasets require parallel utterances, which are costly and difficult to collect, and often unavailable in real-world scenarios. In contrast, in-the-wild data is easier to obtain and captures a more natural emotional expression.

In this work, we address these challenges by proposing TargetSEC, a plug-and-play SEC model that employs an LDM to generate emotion-focused style embeddings for SEC in the arousal dimension. TargetSEC builds on prior attempts that apply LDMs to VC tasks and integrates ideas from SEC's architectures designed for in-the-wild data \cite{promptvcflexiblestylisticvoice, hai2024dreamvoice, Prabhu2023, prabhu2025enhancinginthewildspeechemotion}.

To the best of our knowledge, this is one of the few works to perform SEC on in-the-wild data and the first to integrate an LDM in this context using a continuous arousal representation \cite{EmoCon, Prabhu2023, prabhu2025enhancinginthewildspeechemotion}. We train TargetSEC on the MSP-Podcast V1.10 corpus \cite{MSP} and compare it against recent in-the-wild baselines \cite{EmoCon, Prabhu2023}. Evaluation is conducted using objective evaluation metrics assessing both speech quality and emotion conversion accuracy. Furthermore, we extend the previous evaluation paradigm by including speaker identity preservation and intelligibility evaluation using word error rate (WER).

To summarize, our paper offers three major contributions: 
\begin{enumerate}
    \item We introduce a latent diffusion framework that models emotion-conditioned style embeddings for SEC on in-the-wild data.
    \item Unlike prior text-prompt-conditioned latent diffusion approaches, our method is driven by speaker and continuous emotion embeddings.
    \item  We propose a plug-and-play architecture that enables emotion conversion without modifying the synthesis model backbone. Notably, any continuous or discrete style conditioning (style, emotion, speaker) could be integrated into the architecture without modifying the decoder backbone at inference time.
\end{enumerate}

\section{Methodology}

\begin{figure*}[t]
    \centering
    \includegraphics[width=\textwidth,trim=0 22 0 0,clip]{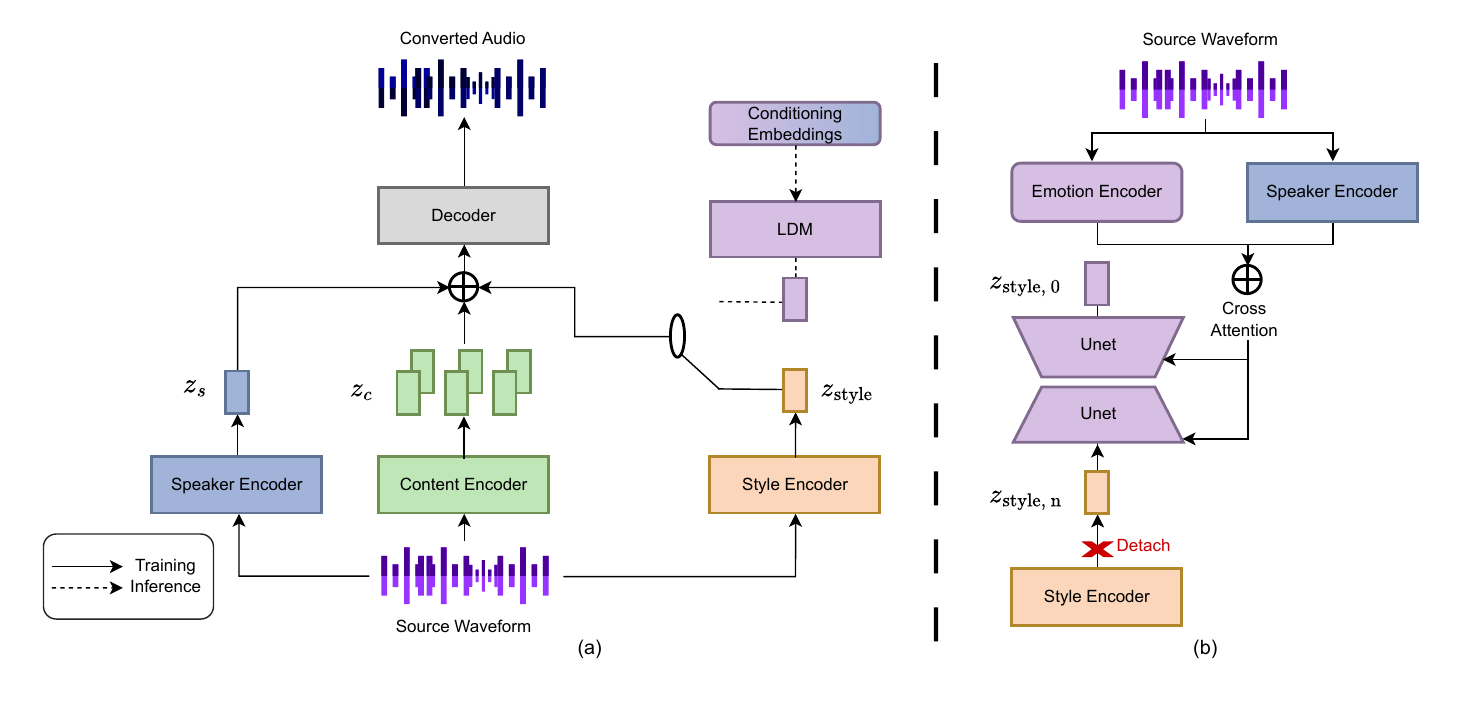}
    \caption{
    Overview of the proposed framework. Subfigure (a) illustrates the architecture of TargetSEC, and (b) shows the training of the LDM.
    Solid arrows denote training paths, while dashed arrows indicate inference-time operations.
    }
    \label{fig:targetsec_architecture}
\end{figure*}
\subsection{Model Architecture}

The overall architecture of TargetSEC can be viewed as an encoder–decoder resynthesis framework, and is illustrated in Figure \ref{fig:targetsec_architecture}. The encoder block consists of three separate encoders, each responsible for disentangling different aspects of the speech: linguistic content, speaker identity, and style. In addition, an emotion encoder provides conditioning for the LDM. The resulting embeddings are concatenated and passed to a HiFi-GAN~V1 decoder \cite{Kong2020}, which reconstructs the target waveform. 

The style encoder is used only during training. At inference time, the style encoder is replaced by the emotion and speaker-conditioned LDM, enabling plug-and-play speech emotion conversion.

Our framework aligns with recent approaches that apply LDMs to speech generation tasks~\cite{AudioLDM, hai2024dreamvoice, promptvcflexiblestylisticvoice}. We adopt a similar encoder-decoder structure to~\cite{promptvcflexiblestylisticvoice} but adapt the methodology for in-the-wild emotion conversion. Specifically, we simplify the conditioning mechanism by replacing text prompts with continuous emotion and speaker embeddings, and we integrate a pre-trained style encoder~\cite{metastylespeechmultispeakeradaptive} to better capture prosodic nuance.

\subsection{Encoder and Representation Disentangling}

\textbf{Content Encoder.}
We use a pretrained HuBERT encoder as the content encoder. Given a time-domain signal $x$, HuBERT extracts a sequence of downsampled representations $E_c(x) = (c_1, \ldots, c_L)$~\cite{Prabhu2023}. These continuous embeddings are quantized using $k$-means clustering, producing a sequence of discrete tokens $u \in \{1, \ldots, K\}$ \cite{polyak2021speechresynthesisdiscretedisentangled}. The tokens are subsequently mapped into a 128-dimensional continuous embedding space~\cite{Prabhu2023}. Finally, we apply window slicing to extract fixed-length segments of $S$ seconds. Since HuBERT operates at a 50\,Hz frame rate, this yields a final content tensor $z_c \in \mathbb{R}^{L \times 128}$, where the sequence length is $L = S \times 50$.

\noindent
\textbf{Speaker Encoder.}
Speaker information is extracted using a pretrained WavLM-based speaker verification model \cite{Speaker} $E_s(x)$, which outputs a global 512-dimensional d-vector ${z}_s$ for each utterance \cite{Prabhu2023}. This vector is broadcast across the
$L$ frames and concatenated with the content embeddings, yielding a combined representation $z_T = (z_c, z_s)$.

\noindent
\textbf{Style Encoder.}
 We employ the style encoder from~\cite{metastylespeechmultispeakeradaptive}, which outputs a 128-dimensional global style vector $z_{\text{style}}$. Instead of using it as a frozen zero-shot representation, we fine-tune it jointly with the decoder on MSP-Podcast using a reduced learning rate, while keeping the content, speaker, and emotion encoders frozen. The adapted $z_{\text{style}}$ is concatenated with the content and speaker embeddings, yielding $z_T=(z_c,z_s,z_{\text{style}})$, and serves as the target latent style prior for LDM training.

\noindent
\textbf{Emotion Encoder.}
To encode emotional information, we use a pretrained emotion recognition model $E_{\text{SER}}$ that was fine-tuned on the MSP-Podcast (v1.7) dataset \cite{MSP, EmoSER}. The model outputs a 1024-dimensional emotion embedding $z_e$ along with continuous predictions for arousal, valence, and dominance. Following the procedure in \cite{Prabhu2023, EmoCon, prabhu2025enhancinginthewildspeechemotion}, the embedding $z_e$ is used both as conditioning input to the LDM and as part of the evaluation pipeline for SEC performance.  

\subsection{Latent Diffusion Model}
We model the conditional style prior with an LDM, approximating
$p(z_{\text{style}} \mid z_s, z_e)$~\cite{AudioLDM}.
Following~\cite{AudioLDM}, we apply the standard forward process to transform
$z_{\text{style}} \in \mathbb{R}^{128}$ with Gaussian noise. 

\noindent
\textbf{Velocity Parameterization.}
Unlike standard LDMs that predict noise $\epsilon$, we follow \cite{hai2024dreamvoice, lin2024commondiffusionnoiseschedules, hai2023dpmtsediffusionprobabilisticmodel} and predict the \textit{velocity} $v_n$ to improve generation stability. The training objective minimizes the velocity estimation loss:
\begin{equation}
\mathcal{L}_{v} = \mathbb{E}_{n,\, z_{\text{style}},\, \epsilon} 
\left[ 
\lVert v_n - v_\theta(z_{\text{style}, n}, n, z_s, z_e) \rVert_2^2 
\right]
\label{eq:lv}
\end{equation}
where the target velocity is defined as:

\begin{equation}
v_n = \sqrt{\bar{\alpha}_n}\,\epsilon - \sqrt{1-\bar{\alpha}_n}\, z_{\text{style}}.
\label{eq:vdef}
\end{equation}

\noindent
\textbf{Rescaled Classifier-Free Guidance.}
During inference, we employ Classifier-Free Guidance (CFG) to steer the generation toward the target emotion. 
\begin{align}
v_{\text{cfg}} &= v_{\text{unc}} + w (v_{\text{cond}} - v_{\text{unc}}) 
\end{align}
where $v_{\text{cond}}$ and $v_{\text{unc}}$ are the conditional and unconditional predictions, and $w$ is the guidance scale.

To reduce artifacts common in CFG, we use the guidance rescaling method of~\cite{hai2024dreamvoice, lin2024commondiffusionnoiseschedules} to obtain the final velocity $\tilde{v}$.

\subsection{Loss}
We optimize the generator using a multi-task objective combining adversarial learning, reconstruction, and feature matching. The total generator loss $L_G$ is defined as:

\begin{equation}
\label{eq:total_loss}
    L_G = \sum_{k} ( L_{\text{adv}}(D_k) + \lambda_{\text{fm}} L_{\text{fm}}(D_k) ) + \lambda_{\text{rec}} L_{\text{rec}}.
\end{equation}

where $L_{\text{adv}}$ and $L_{\text{fm}}$ denote the adversarial and feature matching losses, respectively, following~\cite{Kong2020}. $L_{\text{rec}}$ is the $L_1$ distance between the mel-spectrograms of the ground-truth and synthesized waveforms. We set $\lambda_{\text{fm}} = 2$ and $\lambda_{\text{rec}} = 45$ across all experiments~\cite{Prabhu2023, polyak2021speechresynthesisdiscretedisentangled}.

\subsection{Inference}
At inference time, the style encoder is replaced by the LDM. Following prior work \cite{EmoCon, Prabhu2023}, we construct a target emotion embedding $\overline{e}$ by computing the average of the top $20\%$ of training samples associated with the target arousal level. This embedding $\overline{e}$ and the speaker embedding are then used to condition the LDM, generating a style vector aligned with the target emotion.

\section{Experiments}

\subsection{Experimental Setup}

\textbf{Dataset.} We use the MSP-Podcast v1.10 corpus~\cite{MSP}, a large in-the-wild emotional speech dataset containing over 150{,}000 labeled clips ($\approx 230$ hours). We train on the official ``Train`` partition and use the official ``Development`` partition for validation and checkpoint selection, reserving ``Test1`` exclusively for evaluation.

\noindent
\textbf{Training.}
Training is conducted in two phases: (1)  \textit{Backbone Training:}
The encoder-decoder framework is trained to reconstruct audio from ground-truth style embeddings. We employ window slicing on HuBERT embeddings to train on fixed-length $2.5$\,s segments. The content, speaker, and emotion encoders are kept frozen, while the style encoder is fine-tuned jointly with the decoder.
(2) \textit{LDM Training:}
The LDM is subsequently trained to predict these style embeddings conditioned on emotion and speaker representations. We optimize using AdamW with task-specific learning rates. The diffusion process utilizes $N=1000$ steps with a linear noise schedule $\beta \in [10^{-4}, 0.02]$. For inference, we use $100$ steps with a classifier-free guidance scale $w=4$ and a rescaling factor $\phi=0.7$.

\noindent
\textbf{Metrics.}
We evaluate naturalness with Wav2Vec-MOS (WVMOS)~\cite{WVMOS}, emotion conversion with arousal MSE ($L_{mse}$) and MAE ($L_{abs}$) following prior work~\cite{Prabhu2023, EmoCon, prabhu2025enhancinginthewildspeechemotion}, and intelligibility with Whisper-medium WER. To reduce circularity in SER evaluation, we additionally report a cross-model monotonicity score using an independent IEMOCAP-trained emotion recognizer: for each converted sample, we compute ($P(\mathrm{angry}) + P(\mathrm{happy})$), and report its Spearman correlation with the target arousal level. Speaker preservation is measured by ECAPA-TDNN cosine similarity, and subjective quality/emotion transfer is assessed with a human preference test.

\begin{figure*}[t]
    \centering
\includegraphics[width=\textwidth, trim=0 12 0 0,clip]{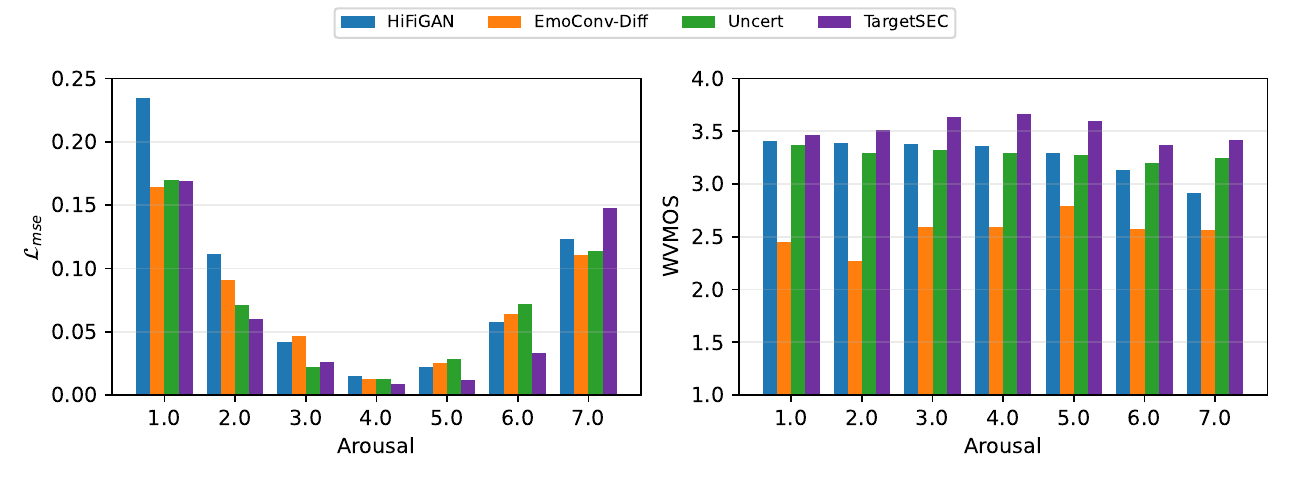}
    \caption{Arousal-wise SER error ($\mathcal{L}_{mse}$ $\downarrow$)  and WVMOS $\uparrow$ for MSP-Podcast Test1.}
    \label{fig:arousal_wvmos_ser}
\end{figure*}

\noindent
\textbf{Model Comparison.}
We compare TargetSEC with the following in-the-wild SEC systems:

\begin{enumerate}
    \item HiFiGAN~\cite{Prabhu2023}: A closely related resynthesis-based SEC baseline with a similar encoder-decoder backbone, using direct emotion-embedding injection instead of latent style diffusion.
    \item EmoConv-Diff~\cite{EmoCon}: A diffusion-based baseline operating on mel-spectrograms.
    \item Uncert~\cite{prabhu2025enhancinginthewildspeechemotion}: A duration-prediction extension of the HiFiGAN baseline.
    \item TargetSEC (Ours): Our proposed framework, replacing direct emotion injection with speaker and emotion-conditioned latent style diffusion.
\end{enumerate}

\subsection{Results}


\begin{table}[t]
\centering
\caption{Objective Test1 performance on MSP-Podcast. DP indicates explicit duration prediction. WVMOS measures naturalness, WER intelligibility, and SER error arousal conversion accuracy.}
\label{tab:main_results}
\small
\setlength{\tabcolsep}{3.0pt}
\renewcommand{\arraystretch}{1.0}
\begin{tabular}{l c c c c c}
\toprule
\textbf{Model} & \textbf{DP} & \textbf{WVMOS}$\uparrow$ & \textbf{WER}$\downarrow$ & $L_{mse}\downarrow$ & $L_{abs}\downarrow$ \\
\midrule
EmoConv-Diff~\cite{EmoCon} & $\times$ & 2.56 & -- & 0.072 & 21\% \\
Uncert~\cite{prabhu2025enhancinginthewildspeechemotion} & $\checkmark$ &  3.30 & -- &  0.069 &  \textbf{20}\% \\
HiFiGAN~\cite{Prabhu2023}$^\dagger$ & $\times$ & 3.27 & 20.7\% & 0.087 & 24.5\% \\
\midrule
TargetSEC (ours) & $\times$ & \textbf{3.52} & \textbf{20.4\%} & \textbf{0.065} & 20.9\% \\
\bottomrule
\end{tabular}

\vspace{0.35em}
\footnotesize{$^\dagger$HiFiGAN results are from our reimplementation under the same MSP-Podcast Test1 split and evaluation pipeline as TargetSEC.}
\end{table}

\textbf{Performance.} TargetSEC achieves the lowest conversion error among all evaluated systems ($L_{mse}=0.065$), outperforming the duration-based Uncert baseline ($L_{mse}=0.069$), despite no explicit temporal modeling. To assess emotion conversion independently of the SER model used for training-target construction, we report the cross-corpus IEMOCAP monotonicity score: TargetSEC reaches $\rho_{\mathrm{IEMO}}=0.450$, versus $0.325$ for the HiFiGAN baseline. As this recognizer was trained on a different corpus and never used in our pipeline, the higher rank correlation provides independent evidence that increasing the target arousal condition produces a monotonic increase in high-arousal posterior beyond the in-domain SER model.

In terms of speech quality, TargetSEC successfully bridges the gap between GAN-based and Diffusion-based approaches. While standard spectrogram diffusion (EmoConv-Diff) suffers a significant degradation in naturalness (WVMOS 2.56), TargetSEC maintains high quality (WVMOS 3.52), exceeding Uncert (3.30). This suggests that performing diffusion in the compact style space avoids the naturalness degradation observed when diffusion operates directly on acoustic representations.

\noindent
\textbf{Influence of Target Arousal.}
Some emotional values are harder to model than others. Related work has shown that SEC models struggle, especially around extreme emotion values. In-the-wild datasets typically exhibit a bias toward mean arousal values (4),  making extreme emotions (e.g., arousal 1 or 7) difficult to model~\cite{EmoCon}. 

Arousal-wise analysis (Figure~\ref{fig:arousal_wvmos_ser}) reveals that TargetSEC outperforms or is on par with all baselines in the mid-to-high arousal range [2.0--6.0], which comprises the majority of in-the-wild speech. Notably, at moderately expressive levels (e.g., Arousal 2 and 6), our model surpasses even the duration-based \textit{Uncert} system.

However, similar to the baselines, TargetSEC's performance degrades at extreme boundaries (Arousal 1 and 7). As noted in~\cite{prabhu2025enhancinginthewildspeechemotion}, extreme emotional states are strongly correlated with speech rate changes (e.g., slowness in boredom, rapidity in anger). Since TargetSEC creates a fixed-duration mapping, it cannot compress or expand the spectrogram to match these temporal shifts, leading to higher error in these edge cases.

Crucially, TargetSEC exhibits superior emotional stability (Figure~\ref{fig:arousal_wvmos_ser}). Unlike HiFiGAN, which suffers quality degradation as arousal increases, TargetSEC maintains consistent naturalness (WVMOS) across the entire emotional spectrum. This suggests that the latent diffusion prior is more robust to high-variance prosody than direct embedding injection.

\noindent
\textbf{Speaker Identity Preservation.}
To verify that TargetSEC preserves speaker identity, we computed the cosine similarity between the embeddings of the source and converted speech using a pretrained ECAPA-TDNN speaker verification model. 
We establish a lower bound by computing the similarity between random speaker pairs ($0.05 \pm 0.09$) and an upper bound by computing the intra-speaker similarity of the ground truth samples ($0.58 \pm 0.17$). 
The relatively low upper bound and high variance of the ground truth reflect the acoustic complexity of the in-the-wild MSP-Podcast corpus. 
TargetSEC achieves a mean similarity of ${0.35} \pm 0.12$, substantially exceeding the random baseline. 
The HiFiGAN baseline attains a comparable score ($0.37 \pm 0.12$), indicating that this partial identity loss reflects the acoustic difficulty of the in-the-wild setting rather than a cost specific to latent-style diffusion. In both cases the similarity remains well above the noise floor, confirming that speaker character is largely preserved despite substantial prosodic modification.

\begin{table}[t]
\centering
\caption{Subjective evaluation ($N=10$) over 30 source--target
conversions per system. Preferences denote TargetSEC's share of
decisive A/B comparisons; ties are excluded.}
\label{tab:subjective}
\small
\setlength{\tabcolsep}{4pt}
\begin{tabular}{lccc}
\toprule
\textbf{System} & \textbf{MOS}$\uparrow$ &
\textbf{Nat. pref.}$\uparrow$ &
\textbf{Emo. pref.}$\uparrow$ \\
\midrule
HiFiGAN   & 2.53 & -- & -- \\
TargetSEC & \textbf{2.98} & \textbf{67.9\%} & \textbf{55.4\%} \\
\bottomrule
\end{tabular}
\end{table}

\subsection{Subjective Evaluation}
We evaluate A1, A4, and A7 with 10 listeners using 5-point naturalness
MOS and pairwise naturalness and target-arousal judgments. Nine of ten
listeners give TargetSEC a higher average MOS (two-sided sign test, $p=0.021$), and it is
chosen as more natural in 67.9\% of decisive comparisons. Target-arousal
preference is 55.4\% overall (61.2\%, 57.3\%, and 48.4\% at A1, A4,
and A7), consistent with Fig.~\ref{fig:arousal_wvmos_ser}.

\subsection{Ablation Study}
To evaluate and validate the impact of our design choices, we conducted an ablation study across four configurations, comparing the proposed LDM against deterministic MLP and assessing the role of speaker conditioning.

\begin{table}[!t]
\centering
\caption{Ablation results comparing the influence of different components.}
\label{tab:ablation}
\small
\setlength{\tabcolsep}{3.5pt}
\begin{tabular}{l c c c c}
\toprule
\textbf{Model Configuration} & \textbf{Diff.} & \textbf{WVMOS} $\uparrow$ & \multicolumn{2}{c}{\textbf{SER Error} $\downarrow$} \\
\cmidrule(lr){4-5}
& & & $L_{mse}$ & $L_{abs}$ \\
\midrule
MLP (Emo Only) & $\times$ & 3.80 & 0.081 & 23.7\% \\
MLP (Speaker + Emo) & $\times$ & \textbf{3.85} & 0.080 & 23.5\% \\
LDM (Emo Only) & $\checkmark$ & 3.63 & 0.067 & 21.6\% \\
\textbf{TargetSEC (Full)} & $\checkmark$ & 3.52 & \textbf{0.065} & \textbf{20.9\%} \\
\bottomrule
\end{tabular}
\end{table}

As shown in Table~\ref{tab:ablation}, we observe a consistent improvement trend: each additional component—switching from regression to diffusion and adding speaker conditioning—progressively reduces the conversion error. While the MLP baselines achieve high naturalness (WVMOS $\approx 3.8$) by reverting to the dataset's mean prosody, they fail to model emotional variety. In contrast, the LDM variants clearly improve conversion accuracy, with the full TargetSEC model achieving the lowest error (20.9\% MAE).

\section{Conclusion and Future Work}
We presented TargetSEC, a plug-and-play SEC framework using latent diffusion to generate emotion-conditioned style embeddings. Operating in compact style space improves in-the-wild SEC while preserving speech quality. Extreme arousal remains challenging due to duration effects; future work will address duration modeling and multidimensional emotion control.

\section{Compliance with Ethical Standards}
\label{sec:ethics}
The subjective listening study involved voluntary adult participants,
who provided informed consent. Responses were collected anonymously,
and no personally identifying or sensitive information was collected.
The study was conducted independently by the author prior to preparation
of this ICASSP submission. No prospective ethics review was obtained,
and retrospective ethics review is not available for the completed study
through the author's institution.

\section{Acknowledgments}
The author thanks the Signal Processing Group at the University of Hamburg
for initial guidance, Jiarui Hai for helpful academic
discussions, and the Artificial Intelligence and Intelligent Systems Group at
the University of Potsdam for assistance in obtaining access to the
MSP-Podcast dataset.

No funding was received for conducting this study. The author has no
relevant financial or nonfinancial interests to disclose.

\bibliographystyle{IEEEbib}
\bibliography{refs}

\end{document}